\documentclass[aos,MSNbibl,seceqn,nameyear,rotating,dvips]{arximspdf}
\usepackage{dcolumn}
\usepackage{graphicx}

\doi{10.1214/12-AOS1030} 
\volume{40}
\issue{4}
\pubyear{2012}
\firstpage{2102}
\lastpage{2127}

\makeatletter

\let\sv@tabnotetext\tabnotetext
\let\sv@tabnotemark@fmt\tabnotemark@fmt
\long\def\legend#1{{\let\tabnote@indent\leavevmode\sv@tabnotetext[]{}{#1}}}

\newcolumntype{d}[1]{D{.}{.}{#1}}

\newcommand{\bbld}{}

\newcommand{\bbbf}[1]{{#1}}

\newtheorem{theorem}{Theorem}[section]
\newtheorem{lemma}{Lemma}[section]

\newproclaim{definition}{Definition}[section]

\makeatother

\begin{document}
\begin{frontmatter}

\title{Generalized fiducial inference for normal linear mixed models}
\runtitle{GFI for normal LMM}

\begin{aug}
\author[A]{\fnms{Jessi} \snm{Cisewski}\corref{}\thanksref{t2}\ead[label=e1]{cisewski@stat.cmu.edu}}
\and
\author[B]{\fnms{Jan} \snm{Hannig}\ead[label=e2]{hannig@email.unc.edu}}
\runauthor{J. Cisewski and J. Hannig}
\affiliation{Carnegie Mellon University and University of North Carolina at~Chapel~Hill}
\address[A]{Department of Statistics\\
Carnegie Mellon University\\
Pittsburgh, Pennsylvania 15213\\
USA\\
\printead{e1}}
\address[B]{Department of Statistics\\
\quad and Operations Research \\
University of North Carolina\\
\quad at Chapel Hill\\
Chapel Hill, North Carolina 27599\\
USA\\
\printead{e2}} 
\end{aug}

\thankstext{t2}{Supported in part by NSF Grants DMS-07-07037,
DMS-10-07543 and DMS-10-16441.}

\received{\smonth{9} \syear{2011}}
\revised{\smonth{7} \syear{2012}}

%
\begin{abstract}
While linear mixed modeling methods are foundational concepts
introduced in any statistical education, adequate general methods for
interval estimation involving models with more than a few variance
components are lacking, especially in the unbalanced setting.
Generalized fiducial inference provides a possible framework that
accommodates this absence of methodology. Under the fabric of
generalized fiducial inference along with sequential Monte Carlo
methods, we present an approach for interval estimation for both
balanced and unbalanced Gaussian linear mixed models. We compare the
proposed method to classical and Bayesian results in the literature in
a simulation study of two-fold nested models and two-factor crossed
designs with an interaction term. The proposed method is found to be
competitive or better when evaluated based on frequentist criteria of
empirical coverage and average length of confidence intervals for small
sample sizes. A MATLAB implementation of the proposed algorithm is
available from the authors.
\end{abstract}

%

\end{frontmatter}

\section{Introduction}

Inference on parameters of normal linear mixed models has an extensive
history; see \citet{KhuriSahai1985} for a survey of variance
component methodology, or Chapter 2 of
\citet{SearleCasellaMcCulloch1992} for a summary. There are many
inference methods for variance components such as ANOVA-based methods
[\citet
{BurdickGraybill1992,HernandezBurdickBirch1992,HernandezBurdick1993,JeyaratnamGraybill1980}],
maximum likelihood estimation (MLE) and restricted maximum likelihood
(REML) [\citet{HartleyRao1967,SearleCasellaMcCulloch1992}] along
with Bayesian methods [\citet
{Gelman2006,GelmanCarlinSternRubin2004,WolfingerKass2000}]. Many of the
ANOVA-based methods become quite complex with complicated models (e.g.,
due to nesting or crossing data structures), and are not guaranteed to
perform adequately when the designs become unbalanced. When the model
design is not balanced, the decomposition of the sum-of-squares for
ANOVA-based methods is not generally unique, chi-squared or
independent. Furthermore, ``exact'' ANOVA-based confidence intervals
are typically for linear combinations of variance components, but not
the \textit{individual} variance components even for simple models
[\citet{BurdickGraybill1992}, pages 67 and 68,
\citet{Jiang2007}]; however, approximate intervals do exist. With
notable optimality properties for point estimation, MLE and REML
methods are less useful when it comes to confidence interval estimation
for small samples because the asymptotic REML-based confidence
intervals tend to have lower than stated empirical coverage
[\citet{Burch2011,BurdickGraybill1992,SearleCasellaMcCulloch1992}].
Bayesian methods, in particular, hierarchical modeling, can be an
effective resolution to complicated models, but the delicate question
of selecting appropriate prior distributions must be addressed.

There are numerous applications of normal linear mixed models related
to topics such as animal breeding studies [\citet
{BurchIyer1997,EHannigIyer2008}], multilevel studies [\citet
{OConnellMcCoach2008}] and
longitudinal studies [\citet{LairdWare1982}]. Many implemented methods
do not go beyond two variance components or are designed for a very
specific setting. We propose a solution based on generalized fiducial
inference that easily allows for inference beyond two variance
components and for the general normal linear mixed model settings.

The proposed generalized fiducial approach is designed specifically for
\textit{interval} data (e.g., due to the measuring instrument's
resolution, rounding for storage on a computer or bid-ask spread in
financial data). There are several reasons to consider interval data.
First, there are many examples where it is critical [or required per
the regulations outlined in \citet{GUM1995}] to incorporate all
known sources of uncertainty
[\citet
{Elster2000,FrenkelKirkup2005,HannigIyerWang2007,LiraWoger1997,Taraldsen2006,Willink2007}].
Second, our simulation results for the normal linear mixed model show
that even when considering this extra source of uncertainty, the
proposed method is competitive or better than classical and Bayesian
methods that assume the data are exact; see Section \ref
{simulationstudy}. [The proposed method is also appropriate for
\textit{noninterval}, or standard, data simply by artificially
discretizing the observation space into a fixed grid of narrow
intervals;
\citet{Hannig2009b} proves that
as the interval width decreases to zero, the generalized fiducial
distribution converges to the generalized fiducial distribution for
exact data.] Finally, on a purely philosophical level, all continuous
data has some degree of uncertainty, as noted previously, due to the
resolution of the measuring instrument or truncation for storage on a
computer. Note that we are not suggesting that all methods should
incorporate this known uncertainty; however, we were able to appeal to
this known uncertainty for the computational aspect of the proposed
method.

A general form of a normal linear mixed model is
%
%
\begin{equation}\label{mlm}
Y = \mathbf X \beta+ \mathbf V Z + \varepsilon,
\end{equation}
where $Y$ is an $n \times1$ vector of data, $\mathbf X$ is a known $n
\times p$ fixed effects design matrix, $\beta$ is a $p \times1$ vector
of unknown fixed effects, $\mathbf V Z = \sum_{i=1}^{r-1} \mathbf V_i
Z_i $, where $Z_i$ is a vector of effects representing each level of
random effect $i$ such that $E(Z_i)=0$ and $\operatorname
{var}(Z_i)=\mathbf G$,
$\mathbf V_i$ is the known design matrix for random effect $i$, and
$\varepsilon$ is an $n \times1$ vector representing the error and
$E(\bbld\varepsilon)=0$ and $\operatorname{var}(\varepsilon)= \mathbf R$
[\citet{Jiang2007}]. Note that there are $r$ total random
components in
this model, and covariance matrices $\mathbf G$ and $\mathbf R$ contain
unknown parameters known as variance components. It is often assumed
that $\mathbf Z$ and $\varepsilon$ are independent and normally
distributed. Additional assumptions on this model for the proposed
method are addressed in Section~\ref{method}.

Except where noted, the notational convention that will be used for
matrices, vectors, and single values will be, respectively, bold and
capital letters for matrices, capital letters for vectors and lowercase
letters for single values (e.g., $\mathbf A$, $A$, $a$).

The focus of this paper is the construction of confidence intervals for
the unknown parameters of~(\ref{mlm}), with emphasis on the variance
components of matrices $\mathbf G$ and $\mathbf R$. In this paper,
inferences are derived from the generalized fiducial distributions of
the unknown parameters, and we propose a sequential Monte Carlo (SMC)
algorithm to obtain these samples. Like a Bayesian posterior, this
procedure produces a distribution on the parameter space, but does so
without assuming a prior distribution. We evaluate the quality of the
simulated generalized fiducial distribution based on the quality of the
confidence intervals---a concept analogous to confidence distributions
[\citet{SchwederHjort2002,XieSinghStrawderman2011}]. We begin by
introducing the two main techniques of the proposed method: generalized
fiducial inference and SMC methods. Then we introduce the proposed
method, and state and prove a theorem concluding the convergence of the
algorithm. To demonstrate small sample performance, we perform a
simulation study on two different types of models (unbalanced two-fold
nested models and two-factor crossed with interaction
models), and include a real-data application for the two-fold nested
model. We finish with concluding remarks. Additional information can be
found in the supplemental document [\citet{CisewskiHannigSupplement2012}].

\subsection{Generalized fiducial inference}
Fiducial inference was introduced by \citet{Fisher1930} to rectify what
he saw as a weakness in the Bayesian philosophy, where a prior
distribution is assumed without\vadjust{\goodbreak} sufficient prior knowledge. While
Fisher made several attempts at justifying his definition of fiducial
inference [Fisher (\citeyear{Fisher1933,Fisher1935a})], it was not fully
developed. Fiducial inference fell into disrepute when it was
discovered that some of the properties Fisher claimed did not hold
[\citet{Lindley1958,Zabell1992}]. Efforts were made to revitalize
fiducial inference by drawing connections to other areas such as
Fraser's structural inference [Fraser
(\citeyear{Fraser1961b,Fraser1961a,Fraser1966,Fraser1968})], and more recently
\citet
{HannigIyerPatterson2006} connect Fisher's fiducial inference to
generalized inference introduced in \citet{TsuiWeerahandi1989} and
\citet{Weerahandi1993}. In this paper, we propose a method for
inference on parameters of normal linear mixed models using the ideas
of generalized fiducial inference.

The main idea of fiducial inference is a transference of randomness
from the model space to the parameter space. A thorough introduction to
generalized fiducial inference can be found in \citet{Hannig2009}, but
here we consider a simple example. Let $y$ be a realization of a random
variable $Y \sim N(\mu, 1)$ [where $N(\mu, 1)$ represents a normal
distribution with unknown mean $\mu$ and standard deviation 1]. The
random variable $Y$ can be represented as $Y=\mu+ Z$ where $Z \sim
N(0,1)$. Given the observed value $y$, the fiducial argument solves
this equation for the unknown parameter $\mu$ to get $\mu= y - Z$;
for example, suppose $y=4.8$, then $\mu= 4.8 - Z$ would suggest $\mu
\sim N(4.8,1)$. While the actual value of $Z$ is unknown, the
distribution of $Z$ is fully known and can be used to frame a
distribution on the unknown parameter $\mu$. This distribution on $\mu
$ is known as the fiducial distribution.

The generalized fiducial recipe starts with a data-generating equation,
also referred to as the \textit{structural equation}, which defines the
relationship between the data and the parameters. Let $Y$ be a random
vector indexed by parameter(s) $\xi\in\Xi$, and then assume $Y$ can
be represented as $Y=G(\xi,{U})$, where $G$ is a jointly measurable
function indicating the structural equation, and $U$ is a random
element with a fully known distribution (void of unknown parameters).
In this paper, the function $G$ will take the form of a normal linear
mixed model, and the random components $U$ will be standard normal
random variables; see equation~(\ref{generalmodel}). Following the
fiducial argument, we define a set-valued function, the inverse image
of $G$, as $Q({\mathbf y},{\mathbf u}) = \{\xi\dvtx{\mathbf y} = G(\xi
,{\mathbf u})\}$,
where ${\mathbf y}$ is the observed data, and ${\mathbf u}$ is an arbitrary
realization of ${U}$. The set-function $Q({\mathbf y},{\mathbf u})$ is then
used to define the fiducial distribution on the parameter space. Since
the distribution of ${U}$ is completely known, independent copies of
$U$, $U^*$, can be generated to produce a random sample of $Q({\mathbf y},
{\mathbf u}^*)$ for the given data ${\mathbf y}$ (where ${\mathbf u}^*$
is a
realization of $U^*$). There are several sources of nonuniqueness in
this framework. In particular, nonuniqueness could occur if $Q$ has
more than one element, if $Q$ is empty, or due to the definition of the
structural equation. Nonuniqueness due to the definition of the
structural equation will not be addressed here as we assume the form of
the model is known (i.e., normal linear mixed model). To resolve the
case when there is more than one element in $Q$, we can define a rule,
call it $V$, for selecting an element of $Q$. Furthermore, since the
parameters $\xi$ are fixed but unknown, there must be some realization
of the random variable $U$ such that ${\mathbf y} = G(\xi,{\mathbf u})$ has
occurred [i.e., $\{Q({\mathbf y},{\mathbf u})\neq\varnothing\}$]. The
generalized fiducial distribution of $\xi$ is defined as
%
%
\begin{equation}\label{fiducialdistribution}
V\bigl(Q\bigl({\mathbf y}, U^*\bigr)\bigr) \mid\bigl\{Q\bigl({\mathbf y},
U^*\bigr)\neq
\varnothing\bigr\}.
\end{equation}
Obtaining a random sample from the generalized fiducial distribution as
defined in~(\ref{fiducialdistribution}) where the structural equation,
$G$, takes the form of a normal linear mixed model is the focus of the
proposed algorithm.

Defining the generalized fiducial distribution as (\ref
{fiducialdistribution}) leads to a potential source of nonuniqueness
due to conditioning on events with zero probability [i.e., if $P(\{
Q(\mathbf y, \mathbf u) \neq\varnothing\})=0$]. This is known as the
Borel Paradox [\citet{CasellaBerger2002}]. Fortunately this can be
resolved by noting that most data has some degree of known uncertainty
due, for example, to the resolution of the instrument collecting the
data or computer storage. Because of this, instead of considering the
value of a datum, an interval around the value can be used [\citet
{Hannig2009b,HannigIyerWang2007}]. For example, suppose the datum
value is $y = 1.632$ meters measuring the height of a woman. If the
resolution of the instrument used to measure the woman is $0.001$ m
(i.e., $1$ mm), then her actual height is between $1.631$~meters and
$1.632$ meters (or between $1.632$ meters and $1.633$ meters, depending
on the practice of the measurer).

By considering interval data, the issue of nonuniqueness due to the
Borel Paradox is resolved since the probability of observing our data
will never be zero since $P(Q((a,b), U^*)\neq\varnothing) \geq P(Y
\in(a,b)) > 0$ where $a<b$ are the endpoints of the interval.

Interval data is not explicitly required for generalized fiducial
inference, but is useful in the proposed setting of normal linear mixed
models. Generalized fiducial inference has a number of other
applications in various settings such as wavelet regression [\citet
{HannigLee2009}], confidence intervals for extremes [\citet
{WandlerHannig2010}], metrology [\citet{HannigIyerWang2007}] and
variance component models [\citet{EHannigIyer2008}], which applies the
generalized fiducial framework to unbalanced normal linear mixed models
with two variance components.

\subsection{Sequential Monte Carlo}
When integrals of interest are very complex or unsolvable by analytical
methods, simulation-based methods can be used. SMC, or particle
filters, is a collection of simulation methods used to sample from an
evolving target distribution (i.e., the distribution of interest)
accomplished by propagating a system of weighted particles through time
or some other index. A solid introduction of and applications to SMC
methods can be found in \citet{DoucetFreitasGordon2001}. There are
many dimensions to the theory and applications of SMC algorithms
[Chopin (\citeyear{Chopin2002,Chopin2004}),
\citet{DelMoralDoucetJasra2006,DoucMoulines2008,KongLiuWong1994,LiuChen1998,LiuWest2001}],
but a simplified introduction is presented below.

Suppose one desires to make inferences about some population based on
data~$Y$. A particle system $\{z^{(J)},w^{(J)}\}$ for $J=1, \ldots, N$
particles is a collection of $N$ weighted random variables (with
weights $w^{(J)}$) such that
\[
\lim_{N \longrightarrow\infty} \frac{\sum_{J=1}^N w^{(J)} \gamma
(z^{(J)})}{\sum_{J=1}^N w^{(J)}} \longrightarrow E_{\pi}\bigl\{
\gamma(z)\bigr\} = \int\gamma(z) \,d\pi(z),
\]
where $\pi$ is the target distribution, and $\gamma$ is some
measurable function, when this expectation exists. Since it is often
difficult to sample directly from the target distribution $\pi$, it
becomes necessary to find some proposal distribution $\widetilde{\pi
}$ to sample the particles. The weights $w^{(J)}$ are determined in
order to re-weight the sampled particles back to the target density
[i.e., $w^{(J)} = \pi(z^{(J)})/\widetilde{\pi}(z^{(J)})$]. This is
the general idea of importance sampling (IS). If the target
distribution is evolving with some time index, an iterative approach to
calculating the weights is performed. This is known as sequential
importance sampling (SIS). Unfortunately, with each time step, more
variation is incorporated into the particle system, and the weights
degenerate, leaving most particles with weights of essentially zero.
The degeneracy is often measured by the effective sample size (ESS),
which is a measure of the distribution of the weights of the particles.
\citet{KongLiuWong1994} presents the ESS as having an inverse relation
with the coefficient of variation of the particle weights, and proved
that this coefficient of variation increases as the time index
increases (i.e., as more data becomes available) in the SIS setting. An
intuitive explanation of ESS can be found in \citet{LiuChen1995}.

SMC builds on ideas of sequential importance sampling (SIS) by
incorporating a resampling step to resolve issues with the degeneracy
of the particle system. Once the ESS for the particle system has
dropped below some designated threshold or at some pre-specified time,
the particle system is resampled, removing inefficient particles with
low weights and replicating the particles with higher weights
[\citet
{LiuChen1995}]. There are various methods for resampling with the most
basic being multinomial resampling, which resamples particles based on
the normalized importance weights; see \citet{DoucCappeMoulines2005}
for a comparison of several resampling methods.

Examples of general SMC algorithms can be found in \citet
{DelMoralDoucetJasra2006} or \citet{JasraStephensHolmes2007}. The main
idea of SMC methods is to iteratively target a sequence of
distributions $\{\pi_t\}_{t \in\mathbf{Z}^+}$, where $\pi_t$ is
often some distribution based on the data available up to time~$t$. The
algorithm comprises three main sections after the initialization step:
sampling, correction and resampling. The sampling step arises at a new
time step $t$ when particles are sampled from some evolving proposal
distribution~$\widetilde{\pi}_t$. The correction step is concerned
with the calculation of the weights and the idea of reweighting the
particles to target the desired distribution at time $t$, $\pi_t$. The
resampling step is performed when the ESS of the particle system falls
below some desired threshold $T$ (e.g., $T=N/2$). The asymptotic
correctness for SMC algorithms can be found in \citet{DoucMoulines2008}.

\section{Method} \label{method}
\subsection{Introduction}
The form of the normal linear mixed model from equation~(\ref{mlm}) is
adapted to work in the generalized fiducial inference setting as
%
%
\begin{equation}\label{generalmodel}
\mathbf Y = \mathbf X \bbld\beta+ \sum_{i=1}^r
\sigma_i \sum_{j=1}^{l_i}
V_{i,j} z_{i,j},
\end{equation}
where $\mathbf X$ is a known $n \times p$ fixed-effects design matrix,
$\beta$ is the $p \times1$ vector of fixed effects, $V_{i,j}$ is the $n
\times1$ design vector for level $j$ of random effect $i$, $l_i$ is the
number of levels per random effect $i$, $\sigma_i^2$ is the variance of
random effect $i$ and the $z_{i,j}$ are independent and identically
distributed standard normal random variables. We will derive a
framework for inference on the unknown parameters ($\beta$~and
$\sigma_i, i=1,\ldots,r$) of this model, which will be applicable to
both the balanced and the unbalanced case. (The design is balanced if
there is an equal number of observations in each level of each effect,
otherwise the design is unbalanced.) In addition, the covariance
matrices for the random components ($\mathbf G$ and $\mathbf R$ above)
are identity matrices with a rank equal to the number of levels of its
corresponding random effect with $l_r = n$. The $\mathbf V_i$, $i = 1,
\ldots, r$, allow for coefficients for the random effects, and
correlation structure can be incorporated into the model by including
additional effects with design matrices that account for the
correlation. We note that these additional assumptions are related to
the implementation of the proposed algorithm, and are not restrictions
of the generalized fiducial framework.

Consider the following example illustrating the connection between
(\ref{mlm}) and~(\ref{generalmodel}) in the case of a one-way random
effects model. A one-way random effects model is conventionally written as
%
%
\begin{equation}\label{oneway}
y_{ij}=\mu+\alpha_i+\varepsilon_{ij},\qquad i=1, \ldots,
a, j=1, \ldots, n_i,
\end{equation}
with unknown mean $\mu$, random effect $\alpha\sim N(0,
\sigma_{\alpha}^2)$ where $\sigma_{\alpha}^2$ is unknown, $a$ is the
number of levels of $\alpha$, $n_i$ is the number of observations in
level $i$ and error terms\vadjust{\goodbreak} $\varepsilon_{ij} \sim N(0,
\sigma_{\varepsilon}^2)$ where $\sigma_{\varepsilon}^2$ is also
unknown, and
$\alpha$ and $\varepsilon$ are independent [\citet{Jiang2007}].
This can be
structured in the form of equation~(\ref{generalmodel}) as
\[
\mathbf Y = \mathbf X \beta+ \sigma_1 \sum
_{j=1}^{l_1} V_{1,j} z_{1,j} +
\sigma_2 \sum_{j=1}^{n}
V_{2,j} z_{2,j},
\]
where $\beta= \mu$ is the overall mean, $\mathbf X = \mathbf1_n$ (an
$n \times1$ vector of ones), $l_1=a$ is the number of levels for the
first random effect and $V_{1,j}$ indicates which observations are in
level $j$ with random effect variance $\sigma_1^2=\sigma_{\alpha
}^2$. The second random effect corresponds to the error, and hence
$V_{2,\cdot}=\mathbf{I}_n$ with $\sigma_{\varepsilon}^2$ as the error
variance component. The $z_{1,\cdot}$ and $z_{2,\cdot}$ are the
i.i.d. standard normal random variables.

The SMC algorithm presented in this section is seeking a weighted
sample of particles $\{Z_{1: t}^{(J)}, W_{1: t}^{(J)}\}_{J=1}^N$ (where
$W_{1: t}^{(J)}$ is the unnormalized importance weight for particle
$Z_{1: t}^{(J)}$) from the generalized fiducial distribution of the
unknown parameters in the normal linear mixed model. Once this sample
of $N$ weighted particles is obtained, inference procedures such as
confidence intervals and parameter estimates can be performed on any of
the unknown parameters or functions of parameters. For example,
parameter estimates can be determined by taking a weighted average of
the particles with the associated (normalized) weights. A $C\%$
confidence interval can be found easily for each parameter by ordering
the particles and finding the particle values $\theta_L$ and $\theta_U$
such that the sum of the normalized weights for the particles
between $\theta_L$ and $\theta_U$ is $C\%$.

\subsection{Algorithm} \label{basicmodelsection}
The algorithm to obtain a generalized fiducial sample for the unknown
parameters of normal linear mixed models is outlined below. As
discussed earlier, the data $Y$ are not observed exactly, but rather
intervals around the data are determined by the level of uncertainty of
the measurement (e.g. due to the resolution of the instrument used).
The structural equation formed as interval data for $t=1,\ldots, n$
with $i=1,\ldots, r$ random effects is
%
%
\begin{equation}\label{fidgeneral}
a_t < Y_t = X_t \bbld\beta+ \sum
_{i=1}^r \sigma_i \sum
_{j=1}^{l_i}v_{i,j,t} z_{i,j} \leq
b_t,
\end{equation}
where the random effect design vector component $v_{i,j,t}$ indicates
the $j$th level of a random effect $i$ for the $t$th element of the
data vector, $X_t $ is the $t$th row of the fixed effect design matrix
$\mathbf X$ and $z_{i,j}$ is a normal random variable for level $j$ of
random effect $i$. Each datum can have one or more random components,
and so we write $Z_{1: t} = (Z_1, \ldots, Z_t)$ with capital letters to
indicate the possible vector nature of each $Z_k$ for $k = 1, \ldots,
t$. In the case that $r>1$, $Z_{1: t}$ is vectorized to be a $\sum
_{i=1}^r l_i \times1$ vector. Also, for notational convenience, $Z_k$
will represent all $z_{i,j}$ for $i=1, \ldots, r$ and $j=1, \ldots,
l_i$ that are not present or shared with any datum less than $k$, which
will always at least include the error effect, denoted $z_{k,r}$. It
will be necessary at times to reference the nonerror random effects,
and they will be denoted $Z_{k, 1: r-1}$, representing all the nonerror
random effects first introduced at time $k$. The goal is to generate a
sample of the $z_{i,j}$ for $i=1, \ldots, r$ and $j=1, \ldots, l_i$
such that $a_t <Y_t \leq b_t$ for $t=1, \ldots, n$.

With a sample of size $n$, the generalized fiducial distribution on the
parameter space can be described as
%
%
\begin{equation}\label{genfiducialdistribution}
V\bigl(Q\bigl((\mathbf a, \mathbf b]_{1: n}, Z_{1: n}\bigr)\bigr) \mid
\bigl\{Q\bigl((\mathbf a, \mathbf b]_{1: n}, Z_{1: n}\bigr) \neq\varnothing
\bigr\},
\end{equation}
where we define the set function $Q$ as the set containing the values
of the parameters that satisfy equation~(\ref{fidgeneral}), given the
data and random component $\mathbf Z$.
Generating a sample from~(\ref{genfiducialdistribution}) is equivalent
to generating the $\mathbf Z$ such that the set $Q$ is nonempty, and
this results in a target distribution at time $t$ written as
%
%
\begin{equation}\label{fulltargetdensity}\quad
\pi_{1: t}\bigl(Z_{1: t} \mid(\mathbf{a},\mathbf{b}]_{1: t}\bigr)=
\pi_{1: t}(Z_{1: t}) \propto\exp\bigl( - \bigl(Z_{1: t}^T
\cdot Z_{1: t}\bigr)/2\bigr) \cdot{\mathbf I_{\mathbf C_t}}(Z_{1: t}),
\end{equation}
where ${\mathbf I_{\mathbf C_t}}(\cdot)$ is an indicator random
variable for the set $\mathbf C_t$, where $\mathbf C_t$ is the set of
$Z_{1: t}$ such that $Q_t$ is not empty. This is equivalent to $\mathbf
C_t = \{Z_{1: t}\dvtx
\exists \beta, \sigma_i$ so that
$a_k < X_k \bbld\beta+ \sum_{i=1}^r \sigma_i
\sum_{j=1}^{l_i}v_{i,j,k} z_{i,j} \leq b_k, k = 1, \ldots, t\}$.
The restriction that $Q_t^{(J)}$ is nonempty can be translated into
truncating the possible values of the particle corresponding to the
error random effect to the interval defined by
\begin{eqnarray*}
&&
m_{t}\bigl(Z_{1: t-1}^{(J)},  Z_{t, 1: r-1}^{(J)}
\bigr)
\\
&&\qquad= \min\biggl(\frac{ a_{t}^{(J)} - ( X_t \beta+ \sum_{i=1}^{r-1} \sigma
_i \sum_{j=1}^{l_i} v_{i,j,t} z_{i,j}^{(J)}
)}{\sigma_r}, (\beta, \sigma) \in
Q_{t-1}^{(J)} \biggr),
\\
&&
M_{t}\bigl(Z_{1: t-1}^{(J)}, Z_{t, 1: r-1}^{(J)}
\bigr)
\\
&&\qquad=\max\biggl(\frac{ b_{t}^{(J)} - (X_t \beta+ \sum_{i=1}^{r-1}
\sigma_i \sum_{j=1}^{l_i} v_{i,j,t} z_{i,j}^{(J)} )}{\sigma_r}, (\beta,
\sigma) \in
Q_{t-1}^{(J)} \biggr).
\end{eqnarray*}
That is, $m_{t} ( Z_{1: t-1}^{(J)}, Z_{t, 1: r-1}^{(J)} )$ and
$M_{t} (Z_{1: t-1}^{(J)}, Z_{t, 1: r-1}^{(J)} )$ are the minimum
and maximum possible values of $z_{t, r}^{(J)}$.

The proposal distribution used is the standard Cauchy distribution due
to improved computational stability of sampling in the tails over the
more natural choice of a standard normal distribution. Specifically,
the $z_{t,r}^{(J)}$ are sampled from a standard Cauchy truncated to
$ (m_{t} (Z_{1: t-1}^{(J)}, Z_{t, 1: r-1}^{(J)} ),
M_{t} (Z_{1: t-1}^{(J)},\break Z_{t, 1: r-1}^{(J)} ) )$ for $t >
p + r$ (i.e., for $t$ greater than the dimension of the parameter
space); otherwise $z_{t,r}^{(J)}$, like $Z_{t,1: r-1}^{(J)}$, is sampled
from a standard normal distribution. The conditional proposal density
at time $t$ for $t > p+r$ is defined as
\begin{eqnarray*}
&&
\widetilde{\pi}_{t \mid1: t-1}\bigl(Z_t \mid Z_{1: t-1},
(\mathbf{a},\mathbf{b}]_{1: t} \bigr) \\
&&\qquad= \widetilde{\pi}_{t \mid1: t-1} (Z_t )
\\
&&\qquad\propto \exp\bigl( - \bigl(Z_{t, 1: r-1}^T \cdot Z_{t, 1: r-1}
\bigr)/2 \bigr) \times\mathbf I_{ (m_t (Z_{1: t-1}, Z_{t, 1: r-1} ),
M_t (Z_{1: t-1}, Z_{t, 1: r-1} ) )} (z_{t,r} )
\\
&&\qquad\quad{} \times\bigl[ \bigl(1+z_{t,r}^2 \bigr) \bigl(F
\bigl(M_t (Z_{1: t-1}, Z_{t, 1: r-1} ) \bigr)-F
\bigl(m_t (Z_{1: t-1}, Z_{t, 1: r-1} ) \bigr) \bigr)
\bigr]^{-1},
\end{eqnarray*}
where $F$ is the standard Cauchy cumulative distribution function. Then
the full proposal density at time $t$ is
%
%
\begin{equation}\label{fullsamplingdensity}\qquad
\widetilde{\pi}_{1: t} \bigl(Z_{1: t} \mid(\mathbf{a},\mathbf{b}]_{1: t}
\bigr) = \widetilde{\pi}_{1: t} (Z_{1: t} ) = \widetilde{
\pi}_1 \prod_{i=2}^t
\widetilde{\pi}_{i \mid1: i-1} \bigl(Z_i \mid Z_{1: i-1},
(\mathbf{a},\mathbf{b}]_{1: i} \bigr).
\end{equation}

The weights are defined as the ratio of the full joint target density
to the full joint proposal density at time t. More specifically, the
weights are derived as
%
%
\begin{equation}\label{weights}
W_{1: t} = \pi_{1: t}/\widetilde{\pi}_{1: t}.
\end{equation}
The resulting sequential updating factor to $W_{1: t-1}$ is $W_t = \exp
(-z_{t,r}^2/2) (1+z_{t,r}^2) (F(M_{t}(Z_{1: t-1}, Z_{t, 1: r-1}) ) -
F(m_{t}(Z_{1: t-1}, Z_{t, 1: r-1}) ))$ where $F$ is the standard Cauchy
cumulative distribution functions. More details regarding the
derivation of these weights can be found in Appendix~\ref{prooftheorem1}.

Standard SMC resampling finds particles according to the distribution
of weights at a given time step, copies the particle and then assigns
the resampled particles equal weight. By copying particles in this
setting, we would not end up with an appropriate distribution on the
parameter space. Intuitively, this is because after each time step,
each particle implies the set, or geometrically the polyhedron, of
possible values of the unknown parameters given the generated
$Z^{(J)}_{1: t}$.

If the particles are simply copied, the distribution of polyhedrons
will be concentrated in a few areas due to particles with initially
higher weight, and will not be able to move from those regions because
subsequent particles would continue to define subsets of the copied
polyhedrons, as outlined in the algorithm presented above. Hence rather
than copy the selected particles exactly, we alter them in a certain
way in order to retain properties of heavy particles, while still
allowing for an appropriate sample of $\{Q_t^{(J)}\}_{J=1}^N$. It can
be thought of as a Gibbs-sampling step in a noncoordinate direction
determined by the selected particle. The precise mathematics of this
step can be found in Appendix~\ref{appresampling}.

The proposed algorithm targets the generalized fiducial distribution of
the unknown parameters of a normal linear mixed model of (\ref
{genfiducialdistribution}) displayed in~(\ref{fulltargetdensity}). The
following theorem confirms that the weighted particle system from the
proposed algorithm achieves this as the sample size $N$ approaches
infinity. The proof is in Appendix~\ref{prooftheorem1}.
%
%
\begin{theorem} \label{theorem1} Given a weighted sample $\{
Z_{1: n}^{(J)},W_{1: n}^{(J)}\}^N_{j=1}$ obtained using the algorithm
presented above targeting~(\ref{fulltargetdensity}), then for any
bounded, measurable function $f$ as $N \longrightarrow\infty$,
\[
\Biggl(\sum_{I=1}^N W_{1: n}^{(I)}
\Biggr)^{-1} \sum_{J=1}^N f
\bigl(Z_{1: n}^{(J)}\bigr) W_{1: n}^{(J)}
\stackrel{P} {\longrightarrow} \int f(Z_{1: n}) \pi_{1: n} =
\pi_{1: n} f(Z_{1: n}).
\]
\end{theorem}

This result holds for slightly weaker conditions, which are outlined in
Appendix~\ref{prooftheorem1}. When the data are i.i.d. (e.g., when the
error effect is the only random component), the confidence intervals
based on the generalized fiducial distribution are asymptotically
correct [\citet{Hannig2009b}]. When the data are not i.i.d., previous
experience and simulation results suggest that the generalized fiducial
method presented above still leads to asymptotically correct inference
as the sample size $n$ increases; see the supplemental document
[\citet{CisewskiHannigSupplement2012}] for a short simulation study
investigating asymptotic properties of the proposed method and
algorithm. The asymptotic exactness of generalized fiducial intervals
for two-component normal linear mixed models was established in
\citet{EHannigIyer2008}; asymptotic exactness of generalized fiducial
intervals for normal linear mixed models is a topic of future research.

\section{Simulation study and applications} \label{simulationstudy}
This simulation study has two parts. In the first part, we consider the
unbalanced two-fold nested model with model designs selected from
\citet
{HernandezBurdickBirch1992}. In the second part, we use the unbalanced
two-factor crossed design with an interaction term with designs
selected from \citet{HernandezBurdick1993}; both sets of designs
include varying levels of imbalance. In addition to the classical,
ANOVA-based methods proposed in \citet{HernandezBurdickBirch1992} and
\citet{HernandezBurdick1993}, we compare the performance of our method
to the $h$-likelihood approach of \citet{LeeNelder1996}, and a Bayesian
method proposed in \citet{Gelman2006}. The purpose of this study
is to
compare the small-sample performance of the proposed method with
current methods using models with varying levels of imbalance. The
methods were compared using frequentist repeated sampling properties.
Specifically, performance will be compared based on empirical coverage
of confidence intervals and average confidence interval length. It is
understood that the selection of a prior distribution influences the
behavior of the posterior; the priors were selected based on\vadjust{\goodbreak}
recommendations in the literature for normal linear mixed models as
noted above. While Bayesian methods do not necessarily maintain
frequentist properties, many practitioners interpret results from
Bayesian analyses as approximately frequentist (i.e., they expect
repeated-sampling properties to approximately hold) due to the
Bernstein--von Mises theorem [\citet{LeCam1986,Vaart2007}], and so
performing well in a frequentist sense has appeal. There are a number
of examples investigating frequentist performance of Bayesian
methodology such as Diaconis and Freedman
(\citeyear{DiaconisFreedman1986b,DiaconisFreedman1986a}),
\citet{GhosalGhoshvVaart2000} and
\citet{MossmanBerger2001}.

It is important to note that the proposed method is not restricted to
the model designs selected for this study, and can be applied to any
normal linear mixed model that satisfies the assumptions from previous
sections, while the included ANOVA methods were developed specifically
for the model designs used in this study. A more efficient algorithm
than the proposed method may be possible for specific model designs,
but one of our goals was to present a mode of inference for any normal
linear mixed model design.

As presented below, the proposed method tends to be conservative with
comparable or shorter intervals than the nonfiducial methods used in
the study.

\subsection{Unbalanced two-fold nested model} For the first part of
the simulation study, we consider the unbalanced two-fold nested linear model
%
%
\begin{equation}\label{model1}
y_{ijk} = \mu+ \alpha_i + \beta_{ij} +
\varepsilon_{ijk}
\end{equation}
for $i=1, \ldots, I$, $j=1, \ldots, J_i$, and $k=1, \ldots, K_{ij}$,
where $\mu$ is an unknown constant and $\alpha_i \sim N(0, \sigma
_{\alpha}^2)$, $\beta_{ij} \sim N(0, \sigma_{\beta}^2)$ and
$\varepsilon_{ijk} \sim N(0, \sigma_{\varepsilon}^2)$.

%
\begin{table}[b]
\caption{Model designs used in the two-fold nested model of
(\protect\ref{model1})}\label{simdesign1}
\begin{tabular*}{\tablewidth}{@{\extracolsep{\fill}}lccccccc@{}}
\hline
\textbf{Design} & $\bolds{\phi_1}$ & $\bolds{\phi_2}$ & $\bolds{\phi}$
& $\bolds{I}$ & $\bolds{J_i}$ & $\bolds{K_{ij}}$ & $\bolds{n}$\\
\hline
MI-1 & 0.9000 & 0.8889 & 0.8090 & 5 & 2,1,1,1,1 & 4,4,2,2,2,2 & 16\\
MI-2 & 0.7778 & 0.7337 & 0.6076 & 3 & 4,2,1 & 1,5,5,5,1,5,1 & 23\\
MI-3 & 1.0000 & 1.0000 & 1.0000 & 3 & 3,3,3 & 2,2,2,2,2,2,2,2,2 & 18\\
MI-4 & 0.4444 & 1.0000 & 0.4444 & 6 & 1,1,1,1,1,7 & 2,2,2,2,2,2 & 24\\
&&&&&& 2,2,2,2,2,2 &\\
MI-5 & 1.0000 & 0.4444 & 0.4444 & 3 & 2,2,2 & 1,1,1,1,1,7 & 12\\
\hline
\end{tabular*}
\legend{Note: $\phi_1$ and $\phi_2$ reflect the degree of imbalance
due to $J_i$ and $K_{ij}$, respectively, and $\phi$ is an overall
measure of imbalance. See~(\ref{model1}) for definitions of $I$, $J_i$
and $K_{ij}$; note sample size $(n) = \sum_i \sum_j
K_{ij}$.}
\end{table}

Table~\ref{simdesign1} displays the model designs used in this part of
the simulation study. Five model designs of \citet
{HernandezBurdickBirch1992} were selected to cover different levels of
imbalance both in the number of nested groups ($J_i$)\vadjust{\goodbreak} and the number of
observations within each group ($K_{ij}$). The parameters $\phi_1$ and
$\phi_2$ reflect the degree of imbalance due to $J_i$ and $K_{ij}$,
respectively. The measures of imbalance listed is based on methods
presented in \citet{Khuri1987} where values range from 0 to 1, and
smaller values suggest a greater degree of imbalance. The parameters'
values used in this part of the study are $\mu= 0$, and the following
combinations\vspace*{1pt} of $(\sigma_{\alpha}^2, \sigma_{\beta}^2,
\sigma_{\varepsilon}^2)$: $\mbox{PI-1} = (0.2, 0.1, 0.7)$,
$\mbox{PI-2} = (0.4, 0.3, 0.3)$, $\mbox{PI-3} = (0.2, 0.7, 0.1)$,
$\mbox{PI-4} = (25, 4, 16)$ and $\mbox{PI-5} = (1, 1, 1)$.

For each model and parameter design combination, 2000 independent data
sets were generated, and 5000 particles were simulated for the proposed
method. \citet{HernandezBurdickBirch1992} present\vspace*{1pt} two methods for
determining confidence intervals for $\sigma_{\alpha}^2$, and
three\vspace*{1pt}
methods for confidence intervals on $\sigma_{\beta}^2$. One of the
methods is based on the confidence interval construction presented in
\citet{TingBurdickGraybillFranklinJeyaratnamLu1990} for balanced
designs (denoted \bbbf{TYPEI}). The other method invokes unweighted sum
of squares and is denoted \bbbf{USS}. We do not consider the
third method presented in \citet{HernandezBurdickBirch1992} for
confidence intervals on $\sigma_{\beta}^2$ because there is not an
analogous method for $\sigma_{\alpha}^2$. We note that for unbalanced
designs, the decomposition of the sum-of-squares is not unique and the
desired distributional properties (independence and chi-squared) do not
generally hold.

The $h$-likelihood approach of \citet{LeeNelder1996} was implemented
using the R package \textit{hglm}, and the results will be referenced as
\bbbf{HLMM}. The $h$-likelihood methodology is an approach for extending
the likelihood in the case of additional, unobserved, random components
[\citet{LeeNelderPawitan2006}]. In the R package \textit{hglm},
inference on the variance components is performed on the $\log$ scale.
This package was selected because it allows for multiple random effects
terms, and it includes standard errors on the estimates of variance components.

A Bayesian method is also considered for comparison. Bayesian
hierarchical models provide a means of constructing confidence
intervals for random-effects models. Part of the art of the Bayesian
methodology is in selecting appropriate prior distributions for the
unknown parameters. For inference on the unknown variance component
parameters when there is no prior information available (i.e., when
seeking a noninformative prior), \citet{Gelman2006} recommends employing
a uniform prior distribution on the standard deviation parameters when
there are a sufficient number of groups (at least 5); otherwise, a
half-$t$ distribution is suggested. Per the recommendation of
\citet{Gelman2006}, both uniform and half-$t$ priors are considered
(denoted $\bbbf{\mathrm{BAY1}_{1.5}}$ and $\bbbf{\mathrm{BAY1}_3}$, and
$\bbbf{\mathrm{BAY2}_{1.5}}$ and $\bbbf{\mathrm{BAY2}_3}$, respectively,
where the subscripts $1.5$ and $3$ specifies the prior scale variable
as explained in Appendix~\ref{bayesappendix}). The R package
\textit{rjags} is used to implement this method; see
Appendix~\ref{bayesappendix} for more details.\looseness=-1

Performance is based on the empirical coverage of $(1-\alpha)100\%$
confidence intervals and average interval length for the parameters of
interest $\theta$. We define\vadjust{\goodbreak} a lower-tailed $(1-\alpha)100\%$
confidence interval on $\theta$ as the interval $(-\infty, U_{\alpha
}]$ such that $P(-\infty< \theta\leq U_{\alpha}) =1 - \alpha$, an
upper-tailed $(1-\alpha)100\%$ confidence interval on $\theta$ as the
interval $[L_{\alpha}, \infty)$ such that $P(L_{\alpha} \leq\theta
< \infty) = 1 - \alpha$ and a two-sided equal-tailed confidence
interval on $\theta$ as the interval $[L_{\alpha/2}, U_{\alpha/2}]$
such that $P(L_{\alpha/2} \leq\theta\leq U_{\alpha/2})= 1 - \alpha
$. Based on the normal approximation to the binomial distribution, we
will consider empirical coverage between $94\%$ and $96\%$ appropriate
for 95\% two-sided confidence intervals.

\begin{sidewaysfigure}
\begin{center}
\includegraphics{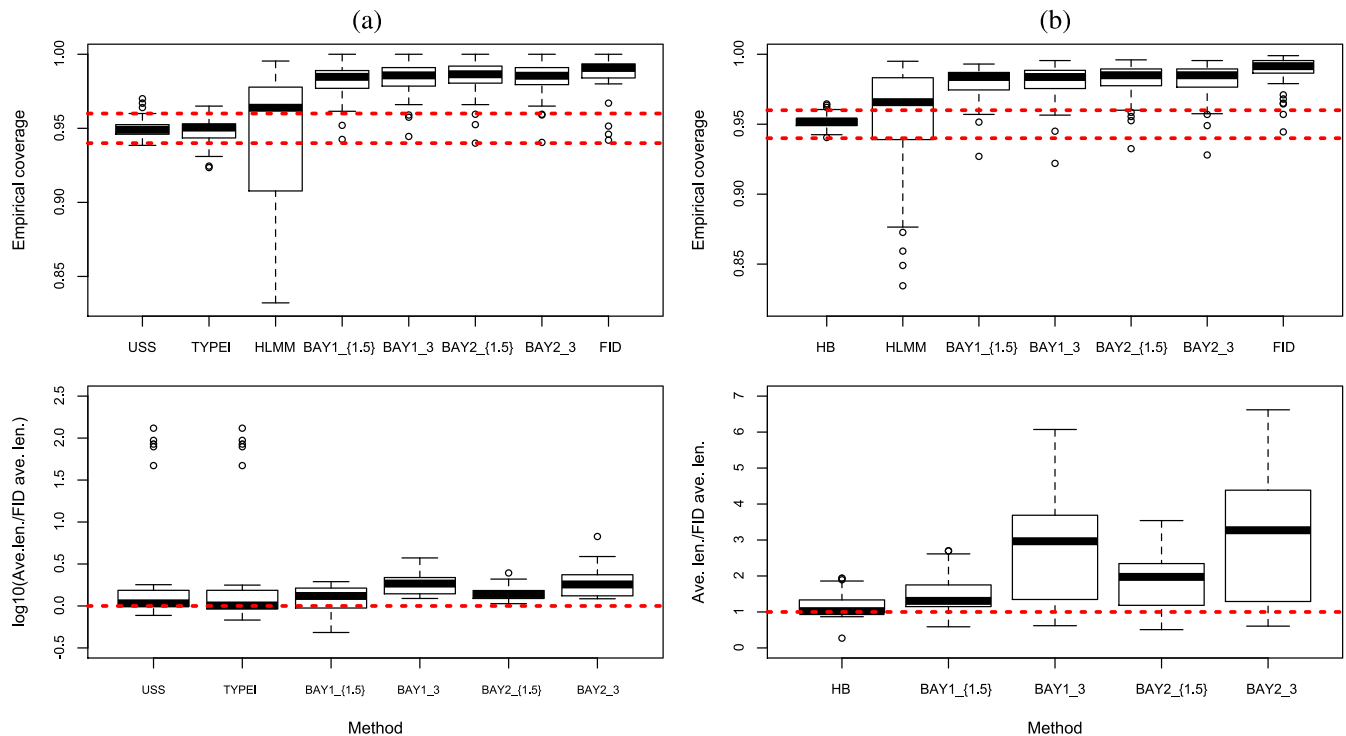}
\end{center}
\caption{Summary\vspace*{-1pt} of simulation results. \textup{(a)} Displays the combined
results for $\sigma_{\alpha}^2$ and $\sigma_{\beta}^2$ for the two-fold
nested model. \textup{(b)} Displays the combined results for
$\sigma_{\alpha}^2$, $\sigma_{\beta}^2$ and $\sigma_{\alpha\beta}^2$
for the two-factor crossed design with interaction. The first row is
the empirical coverage probabilities for 95\% two-sided confidence
intervals. The second row, first column is the $\mathrm{log}_{10}$ of
the average interval lengths divided by the average interval lengths of
\bbbf{FID}, and the second row, second column is the average interval
lengths divided by the average interval lengths of \bbbf{FID}. Each
value used in a box plot corresponds to a particular model design and
parameter combination for each nonerror variance component. The
average interval lengths for \bbbf{HLMM} were not included in the
displayed results due to their excessive lengths. All the simulation
results are available in the supplemental document
[Cisewski and Hannig
(\citeyear{CisewskiHannigSupplement2012})].} \label{boxplotsim}
\end{sidewaysfigure}

A summary of the two-sided 95\% confidence interval results are
displayed in Figure~\ref{boxplotsim}. In addition, the supplemental
document [\citet{CisewskiHannigSupplement2012}] includes figures with
the results summarized by parameter along with the complete raw data
results. Average interval lengths are not included in the displayed
results for \bbbf{HLMM} because the excessive lengths would skew the
scale of the plots; however, the average lengths are displayed in the
raw data results in the supplemental document [\citet
{CisewskiHannigSupplement2012}].

%
%

\bbbf{HLMM} has empirical coverage below the stated level, and the
overall longest average interval lengths. $\bbbf{\mathrm{BAY1}_{1.5}}$,
$\bbbf{\mathrm{BAY1}_3}$, $\bbbf{\mathrm{BAY2}_{1.5}}$ and
$\bbbf{\mathrm{BAY2}_3}$ tend to be conservative with the longest
average interval lengths after \bbbf{HLMM}. \bbbf{USS} and \bbbf{TYPEI}
maintain the stated coverage well, and \bbbf{FID} tends to be
conservative. Even though \bbbf{FID} is conservative, its average
interval lengths are comparable or shorter than those of \bbbf{USS} and
\bbbf{TYPEI}. However, the average interval lengths of \bbbf{USS} and
\bbbf{TYPEI} for $\sigma_{\beta}^2$ are\vspace*{1pt} surprisingly wider than
$\bbbf{\mathrm{BAY1}_{1.5}}$, $\bbbf{\mathrm{BAY1}_3}$,
$\bbbf{\mathrm{BAY2}_{1.5}}$, $\bbbf{\mathrm{BAY2}_3}$ and \bbbf{FID}
for MI-1, as revealed in the plot of the average lengths in Figure
\ref{boxplotsim}(a). This is due to the model design; specifically, the
derivation of the confidence intervals results in the degrees of
freedom of 1 for the nested factor (calculated as $\sum_{i=1}^IJ_i - I
= 1$). $\bbbf{\mathrm{BAY1}_{1.5}}$, $\bbbf{\mathrm{BAY1}_3}$,
$\bbbf{\mathrm{BAY2}_{1.5}}$, $\bbbf{\mathrm{BAY2}_3}$ and \bbbf{FID} do
not appear to have this issue. Upper and lower one-sided confidence
interval results for $\sigma_{\alpha}^2$ and $\sigma_{\beta}^2$ for
\bbbf{USS} and \bbbf{TYPEI} tend to stay within the stated level of
coverage, and $\bbbf{\mathrm{BAY1}_{1.5}}$, $\bbbf{\mathrm{BAY1}_3}$,
$\bbbf{\mathrm{BAY2}_{1.5}}$, $\bbbf{\mathrm{BAY2}_3}$ and \bbbf{FID}
range from staying within the stated level of coverage to very
conservative.

The proposed method, while maintaining conservative coverage, has
average interval lengths that are competitive or better than the other
methods used in this part of the study. The proposed method offers an
easily generalizable framework and provides intervals for fixed effects
and the error variance component, unlike the methods presented in
\citet
{HernandezBurdickBirch1992}. While the conservative coverage for the
Bayesian methods can be deemed acceptable, their average interval
lengths tend to be wider than the proposed method.

\subsubsection{Application 1} \label{sectionapp1}
In addition to the simulation study, we consider the application of
model~(\ref{model1}) presented in \citet{HernandezBurdickBirch1992}
concerning the blood pH of female mice offspring. Fifteen dams were
mated with 2 or 3 sires, where each sire was only used for one dam
(i.e., 37 sires were\vadjust{\goodbreak} used in the experiment), and the purpose of the
study was to determine if the variability in the blood pH of the female
offspring is, in part, due to the variability in the mother. There is
imbalance in the data due to the number of sires mated with each dam (2
or 3); also note the natural imbalance in the data resulting from the
number of female offspring.

\begin{figure}

\includegraphics{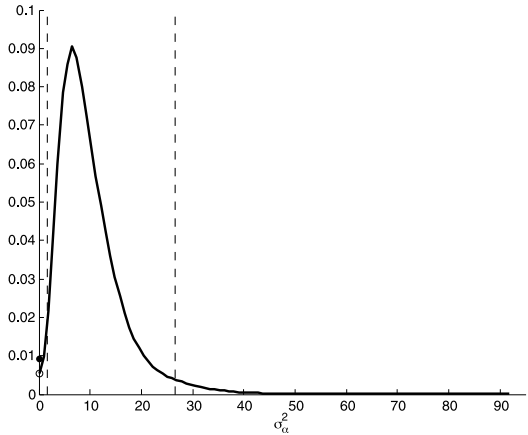}

\caption{The generalized fiducial distribution of 5000 generated
particles for $\sigma_{\alpha}^2$ using the real data described in
Section \protect\ref{sectionapp1}. The solid line is the normal kernel
density estimate of the distribution with a point mass at zero. The
two vertical dashed lines represent the lower and upper bounds for 95\%
confidence intervals based on the weights of the generated particles.}
\label{sampleexample}
\end{figure}

The 95\% confidence intervals based on the real data are presented in
Table~\ref{app1}. An example of the generalized fiducial distribution
for $\sigma_{\alpha}^2$ is displayed in Figure~\ref{sampleexample}.
This highlights one of the advantages of the proposed method over
classical methods (and shared with Bayesian methods), which is a
distribution on the parameter space allowing for inferences similar to
those made using Bayesian posterior distributions.

%
\begin{table}
\caption{Two-fold nested model: real data example}
\label{app1}
\begin{tabular*}{\tablewidth}{@{\extracolsep{\fill}}lcccc@{}}
\hline
\textbf{Var. comp.} & \textbf{Method} & \textbf{95\% 2-sided CI}
& \textbf{2-sided/ave. len.} &
\textbf{Upper/lower} \\
\hline
$\sigma_{\alpha}^2$ & \bbbf{USS} & (2.30, 28.56) & 0.953$/$25.1 &
0.949$/$0.958\\
& \bbbf{TYPEI} & (1.94, 26.23) & 0.950$/$25.2 & 0.949$/$0.957 \\
& $\bbbf{\mathrm{BAY1}_{1.5}}$ & (1.73, 30.72) & 0.955$/$29.2 &
0.948$/$0.959 \\
& $\bbbf{\mathrm{BAY1}_{3}}$ & (1.51, 30.21) & 0.956$/$37.2 &
0.948$/$0.962\\
& $\bbbf{\mathrm{BAY2}_{1.5}}$ & (1.56, 30.02) & 0.955$/$27.3 &
0.948$/$0.960 \\
& $\bbbf{\mathrm{BAY2}_{3}}$ & (1.76, 30.04) & 0.955$/$27.5 & 0.950$/$0.961
\\
& \bbbf{FID} & (1.53, 26.67) & 0.947$/$24.5 & 0.958$/$0.947 \\ 
[6pt]
$\sigma_{\beta}^2$ & \bbbf{USS} & (0.00, 11.56) & 0.961$/$10.9 &
0.952$/$0.952\\
& \bbbf{TYPEI} & (0.00, 11.26) & 0.964$/$12.4 & 0.953$/$0.953 \\
& $\bbbf{\mathrm{BAY1}_{1.5}}$ & (0.17, 11.81) & 0.976$/$11.2 &
0.956$/$0.994 \\
& $\bbbf{\mathrm{BAY1}_{3}}$ & (0.04, 12.55) & 0.980$/$11.3 & 0.959$/$0.996
\\
& $\bbbf{\mathrm{BAY2}_{1.5}}$ & (0.01, 12.49) & 0.982$/$11.3 &
0.958$/$0.994 \\
& $\bbbf{\mathrm{BAY2}_{3}}$& (0.01, 11.92) & 0.983$/$11.3 & 0.958$/$0.995
\\
& \bbbf{FID} & (0.19, 10.54) & 0.974$/$10.7 & 0.951$/$0.986\\
\hline
\end{tabular*}
\legend{Note: The 95\% intervals are based on the actual data while
the remaining information are the empirical results from 2000
independently generated data set using the REML estimates for each
parameter. The results are the empirical coverage and average interval
lengths of 95\% confidence
intervals.}
\end{table}

In order to evaluate the empirical coverage of the proposed method, we
perform a simulation study using the REML estimates for all the
parameters ($\mu= 44.92$, $\sigma_{\alpha}^2=8.90$,
$\sigma_{\beta}^2=2.65$ and $\sigma_{\varepsilon}^2=24.81$).
Simulating 2000 independent data sets with the noted parameter values,
we find the empirical coverage using \bbbf{USS}, \bbbf{TYPEI},
$\bbbf{\mathrm{BAY1}_{1.5}}$, $\bbbf{\mathrm{BAY1}_3}$,
$\bbbf{\mathrm{BAY2}_{1.5}}$, $\bbbf{\mathrm{BAY2}_3}$ and \bbbf{FID},
and the average lengths of the two-sided intervals. The results of the
simulation study are also found in Table~\ref{app1}.

%

The confidence interval coverage and average lengths for
$\sigma_{\alpha}^2$ are comparable for all methods with
$\bbbf{\mathrm{BAY1}_3}$ having the longest average interval length.
For $\sigma_{\beta}^2$, $\bbbf{\mathrm{BAY1}_{1.5}}$,
$\bbbf{\mathrm{BAY1}_3}$, $\bbbf{\mathrm{BAY2}_{1.5}}$,
$\bbbf{\mathrm{BAY2}_3}$ and \bbbf{FID} tend to be more conservative
while \bbbf{USS} and \bbbf{TYPEI} correctly maintain the stated level of
coverage. \bbbf{USS}, \bbbf{TYPEI} and \bbbf{FID} have comparable interval
lengths, and the average lengths of $\bbbf{\mathrm{BAY1}_{1.5}}$,
$\bbbf{\mathrm{BAY1}_3}$, $\bbbf{\mathrm{BAY2}_{1.5}}$,
$\bbbf{\mathrm{BAY2}_3}$ are slightly longer.

\subsection{Unbalanced two-factor crossed design with interaction} In
this part of the simulation study, we consider the unbalanced
two-factor crossed designs with interaction written as
%
%
\begin{equation}\label{model2}
Y_{ijk} = \mu+ \alpha_i + \beta_j + (\alpha
\beta)_{ij} + \varepsilon_{ijk}
\end{equation}
for $i=1, \ldots, I$, $j=1, \ldots, J$ and $k=1, \ldots, K_{ij}$,
where $\mu$ is an unknown constant and $\alpha_i \sim N(0, \sigma
_{\alpha}^2)$, $\beta_j \sim N(0, \sigma_{\beta}^2)$, $(\alpha
\beta)_{ij} \sim N(0, \sigma_{\alpha\beta}^2)$ and $\varepsilon_{ijk}
\sim N(0, \sigma_{\varepsilon}^2)$.\vspace*{1pt}

This model is presented in \citet{HernandezBurdick1993}, where the
authors propose a method based on unweighted sum of squares to
construct confidence intervals for $\sigma_{\alpha}^2$, $\sigma_{\beta
}^2$ and $\sigma_{\alpha\beta}^2$. The method\vspace*{1pt} they propose
is based on intervals for balanced designs presented by \citet
{TingBurdickGraybillFranklinJeyaratnamLu1990}. In the simulation study,
this method will be called \bbbf{HB}. As with~(\ref{model1}),
\bbbf{HLMM} from \citet{LeeNelder1996}, and
$\bbbf{\mathrm{BAY1}_{1.5}}$, $\bbbf{\mathrm{BAY1}_3}$,
$\bbbf{\mathrm{BAY2}_{1.5}}$ and $\bbbf{\mathrm{BAY2}_3}$ from
\citet{Gelman2006} will be used as a comparison.

Table~\ref{simdesign2} displays the model designs used in this part
of the study, and, again, the overall measure of imbalance ($\phi$)
proposed in \citet{Khuri1987} is displayed for each design. The
parameters values used in this part of the study are $\mu= 0$, and the
following combinations of $(\sigma_{\alpha}^2, \sigma_{\beta}^2,
\sigma_{\alpha\beta}^2, \sigma_{\varepsilon}^2)$: $\mbox{PII-1} = (0.1,
0.5, 0.1, 0.3)$, $\mbox{PII-2} = (0.1, 0.3, 0.1, 0.5)$, $\mbox
{PII-3} =
(0.1, 0.1, 0.3, 0.5)$, $\mbox{PII-4} = (0.1, 0.1, 0.5, 0.3)$, and
$\mbox{PII-5} =(1, 1, 1, 1)$.

%
\begin{table}
\caption{Model designs used in the two-factor crossed design with
interaction model of (\protect\ref{model2})}
\label{simdesign2}
\begin{tabular*}{\tablewidth}{@{\extracolsep{\fill}}ld{1.4}cccc@{}}
\hline
\textbf{Design} & \multicolumn{1}{c}{$\bolds{\phi}$} & $\bolds{I}$ & $\bolds{J}$
& $\bolds{K_{ij}}$ & $\bolds{n}$\\
\hline
MII-1 & 0.8768 & 4 & 3 & $2, 1, 3/ 2, 1, 1/ 2, 2, 2/ 1, 2, 3$ & 22\\
MII-2 & 0.6667 & 3 & 3 & $4, 1, 1/ 4, 1, 1/ 4, 1, 1$ & 18\\
MII-3 & 0.6667 & 3 & 3 & $4, 4, 4/ 1, 1, 1/ 1, 1, 1$ & 18\\
MII-4 & 0.4011 & 3 & 4 & $8, 1, 1, 1 / 1, 1, 1, 1/ 1, 1, 1, 1$ & 19\\
MII-5 & 0.7619 & 5 & 3 & $1,2,2/5,2,7/2,2,2/2,4,2/3,2,2$ & 40\\
MII-6 & 1.000 & 3 & 3 & $2,2,2/2,2,2/2,2,2$ & 18\\
\hline
\end{tabular*}
\legend{Note: The parameter $\phi$ is an overall measure of imbalance
of the model. See~(\ref{model2}) for definitions of $I$, $J$ and
$K_{ij}$; note sample size $(n) = \sum_i \sum_j
K_{ij}$.}
\end{table}

For each design and set of parameter values, 2000 independent data sets
were generated, and 5000 particles were simulated for the proposed
method. As before, performance is based on the empirical coverage of
$(1-\alpha)100\%$ confidence intervals and average interval length for
the parameter of interest~$\theta$. Based on the normal approximation
to the binomial distribution, we will consider empirical coverage
between $94\%$ and $96\%$ appropriate for 95\% two-sided confidence intervals.

A summary of the two-sided 95\% confidence interval results for $\sigma
_{\alpha}^2$, $\sigma_{\beta}^2$ and $\sigma_{\alpha\beta}^2$ are
displayed in Figure~\ref{boxplotsim}. In addition, the supplemental
document [\citet{CisewskiHannigSupplement2012}] includes figures with
the results summarized by parameter along with the complete raw data
results. Average interval lengths are not included in the displayed
results for \bbbf{HLMM} because the excessive lengths would skew the
scale of the plots; however, the average lengths are displayed in the
raw data results in the supplemental document
[\citet{CisewskiHannigSupplement2012}].

\bbbf{HLMM} has empirical coverage below the stated level, and the
longest average interval lengths. While \bbbf{HB} maintains the stated
coverage and \bbbf{FID} tends to be more conservative, they have
comparable average interval lengths. $\bbbf{\mathrm{BAY1}_{1.5}}$,
$\bbbf{\mathrm{BAY1}_3}$, $\bbbf{\mathrm{BAY2}_{1.5}}$, and
$\bbbf{\mathrm{BAY2}_3}$ are conservative with the longest average
interval lengths after \bbbf{HLMM} for $\sigma_{\alpha}^2$ and
$\sigma_{\beta}^2$; while still conservative for
$\sigma_{\alpha\beta}^2$, the average interval lengths are shorter than
\bbbf{HB}, but longer than \bbbf{FID}. One-sided confidence interval
results for \bbbf{HB} maintains the stated coverage, and
$\bbbf{\mathrm{BAY1}_{1.5}}$, $\bbbf{\mathrm{BAY1}_3}$,
$\bbbf{\mathrm{BAY2}_{1.5}}$, $\bbbf{\mathrm{BAY2}_3}$ and \bbbf{FID}
tends to be within the stated coverage to very conservative.

\section{Conclusion}
Even with the long history of inference procedures for normal linear
mixed models, a good-performing, unified inference method is lacking.
ANOVA-based methods offer, what tend to be, model-specific solutions.
While Bayesian methods allow for solutions to very complex models,
determining an appropriate prior distribution can be confusing for the
nonstatistician practitioner. In addition, for the models considered in
the simulation study and the prior selected based on recommendations in
the literature, the Bayesian interval lengths were not generally
competitive with the other methods used in the study. The proposed
method allows for confidence interval estimation for all parameters of
balanced and unbalanced normal linear mixed models. In general, our
recommendation is to use the proposed method because of its apparent
robustness to design imbalance, good small sample properties and
flexibility of inference due to a fiducial distribution on the
parameter space. If the design is balanced and only confidence
intervals are desired, an ANOVA-based method would provide a
computationally efficient solution.

It is interesting to note that even though more variation was
incorporated into the data for the proposed method due to its
acknowledgment of known uncertainty using intervals, in the simulation
study, the proposed method tended to have conservative coverage, but
the average interval lengths were comparable or shorter than the other
methods that assumed the data are observed exactly. The currently
implemented algorithm is suitable for 9 or fewer total parameters, but
the method does not limit the number of parameters. A MATLAB
implementation of the proposed algorithm is available on the author's
website at
\texttt{\href{http://www.unc.edu/\textasciitilde hannig/download/LinearMixedModel\_MATLAB.zip}%
{http://www.unc.edu/\textasciitilde hannig/download/}
\href{http://www.unc.edu/\textasciitilde hannig/download/LinearMixedModel\_MATLAB.zip}%
{LinearMixedModel\_MATLAB.zip}}.

%
\begin{appendix}\label{app}

\section{Resampling alteration step} \label{appresampling}
The particle to be resampled is decomposed into an orthogonal
projection onto a certain space (discussed below) and the part
orthogonal to that space, and then the distributional properties that
arise from the decomposition are used to alter the resampled particle.
The alteration step of the proposed algorithm is performed in such a
way that it still solves the system of inequalities of (\ref
{fidgeneral}) up to time $t$ using the following idea (to ease the
notational complexity, we do not include the dependence of the
variables on $t$). Suppose particle $L$ is selected to be resampled
(for an $L$ between 1 and $N$). For each random effect, $e$, let $Y =
{\mathbf X}' {\bbld\beta}' + \sigma{\mathbf V} Z^{(L)}$, where
${\mathbf X}'=[\mathbf X, \{\sum_{j=1}^{l_i} V_{i,j} z_{i,j}^{(L)}\}
_{i \neq e}]$, $\beta'=(\beta,\{\sigma_i\}_{i \neq e})'$, $\sigma
=\sigma_e$ and ${\mathbf V} Z^{(L)}=\sum_{j=1}^{l_e} V_{e,j}
Z_{e,j}^{(L)}$. In order to alter $Z^{(L)}$, we first find the basis
vectors, $\eta$, for the null space $\mathcal N = \operatorname
{null}[-{\mathbf X}', {\mathbf V}] = {\bbld\eta}$ where for matrix
${\mathbf A}=[-{\mathbf X}', {\mathbf V}] $,
$\operatorname{null}({\mathbf A})$ is the set $\{\eta\dvtx {\mathbf
A}\eta= {\mathbf0}\}$, and ${\bbld\eta} = ({\bbld\eta}_1,
{\bbld\eta}_2)^T$ such that (i) ${\mathbf X}' \cdot{\bbld\eta}_1 =
{\mathbf0}$, (ii)~${\mathbf V} \cdot{\bbld\eta}_2 = {\mathbf0}$,
(iii)~${\bbld\eta}_2^T\cdot{\bbld\eta}_2 = {\mathbf I}$ (i.e.,
${\bbld\eta }_2$ is orthonormal) and (iv) $\operatorname{rank}(\eta) =
\operatorname{rank}(\eta_2)$. We perform the following decomposition:
%
%
\begin{equation}\label{decomposition}
Z^{(L)} = \Pi Z^{(L)} + \bigl\|Z^{(L)} - \Pi
Z^{(L)}\bigr\| \cdot\frac{
Z^{(L)}- \Pi Z^{(L)}}{\|Z^{(L)}- \Pi Z^{(L)}\|},
\end{equation}
where\vspace*{1pt} $\Pi Z^{(L)}$ is the projection onto the null space $\mathcal N$
(i.e., $\Pi Z^{(L)} = \eta\cdot\eta^T \cdot Z^{(L)}$), and \mbox{$\|\cdot
\|$} is the $L^2$ norm. Define ${\mathbf C} = \eta_2^T \cdot Z^{(L)}$ (so
that, $\eta_2 \cdot{\mathbf C} = \Pi Z^{(L)}$), ${\mathbf D} = \|
Z^{(L)} -
\Pi Z^{(L)}\|$, and $\tau= \frac{Z^{(L)}- \Pi Z^{(L)}}{\|Z^{(L)}- \Pi
Z^{(L)}\|}$. Then if $Z^{(L)}$ is standard normal, ${\mathbf C} \sim
N_{l_e}(0,{\mathbf I})$ where $l_e$ is the number of levels of random
effect $e$ of $Z^{(L)}$ to be resampled at time $t$, and ${\mathbf D}
\sim
\sqrt{\chi^2_{t-d}}$ where $d=\operatorname{rank}(\mathcal N)$, and
$\mathbf C$ and $\mathbf D$ are independent by design. The alteration
of $Z^{(L)}$ is accomplished by sampling new values of ${\mathbf C}$ and
${\mathbf D}$ (denoted $\widetilde{{\mathbf C}}$ and $\widetilde
{{\mathbf D}}$,
resp.) according to their distributions \textit{determined by the
decomposition} above, and the altered particle is
%
%
\begin{equation}\label{generalresampling}
\widetilde{Z} = \eta_2 \cdot\widetilde{{\mathbf C}} + \widetilde
{{\mathbf D}}\cdot\tau.
\end{equation}
Notice that if $Z^{(L)}$ is a standard normal conditioned on $\mathbf
C_t$, then so is $\widetilde{Z}$, and hence the alteration proposed
still targets the correct distribution and is a Markovian step.

Furthermore, the set $Q_t^{(L)} = \{(\beta', \sigma)\dvtx a_i <
{\mathbf X}'
{\bbld\beta'} + \sigma{\mathbf V} Z^{(L)} \leq b_i, i = 1, \ldots, t\}$
can be adjusted noting that if $(\beta', \sigma)$ solves $a_i <
{\mathbf X}' \beta' + \sigma{\mathbf V} (\eta_2 \cdot{\mathbf C} +
{\mathbf D} \cdot
{\bbld\tau}) \leq b_i$ for $i=1, \ldots, t$, then $(\widetilde{\beta
}, \widetilde{\sigma})$ can be\vspace*{1pt} found such that $a_i < {\mathbf X}'
\widetilde{\beta} + \widetilde{\sigma} {\mathbf V} (\eta_2 \cdot
\widetilde{\mathbf C} + \widetilde{\mathbf D} \cdot{\bbld\tau}) \leq b_i$
for $i=1,\ldots,t$ by\vspace*{1pt} considering $\mathbf X' \beta+ \sigma\mathbf
{V}Z = \mathbf X \beta+ \sigma\mathbf V (\eta_2\cdot\mathbf C
+ \tau\mathbf D ) =
\mathbf X' \widetilde{\beta} + \widetilde{\sigma}\mathbf V
(\eta_2 \cdot\widetilde{\mathbf C} + \tau\widetilde{\mathbf D} )=
\mathbf X' \widetilde{\beta} + \widetilde{\sigma}\mathbf V
\widetilde{Z}$. Examining\vspace*{1pt} the orthogonal parts first, $\sigma\mathbf
V(\tau\mathbf D) = \widetilde{\sigma} \mathbf V (\tau\widetilde
{\mathbf D}) \mbox{ implies } \mathbf V \tau(\sigma\mathbf D
-\widetilde{\sigma} \widetilde{\mathbf D} )=0$ implies
$\widetilde{\sigma} = \sigma(\mathbf D/\widetilde{\mathbf
D} )$. The relation between $\widetilde{\beta}$ and $\beta'$
follows from the remaining portion
%
%
\begin{eqnarray}\label{relation1}
&\displaystyle \mathbf X' \beta' + \sigma\mathbf V (
\eta_2\cdot\mathbf C ) = \mathbf X' \widetilde{\beta} +
\widetilde{\sigma}\mathbf V (\eta_2 \cdot\widetilde{\mathbf C} )
\quad\mbox{implies}&
\nonumber\\[-8pt]\\[-8pt]
&\displaystyle \mathbf X'\bigl(\widetilde{\beta}-\beta' \bigr) +
\sigma\mathbf V \eta_2 (\widetilde{\mathbf C}\cdot\mathbf D/
\widetilde{\mathbf D} - \mathbf C ) = 0.&\nonumber
\end{eqnarray}
Noting by definition $-\mathbf X' \eta_1 + \mathbf V \eta_2
= 0$, then
%
%
\begin{equation} \label{relation2}
-\mathbf X' \sigma\eta_1 (\widetilde{\mathbf C} \cdot
\mathbf D/\widetilde{\mathbf D} - \mathbf{C} ) + \sigma\mathbf V
\eta_2 (\widetilde{\mathbf C} \cdot\mathbf D/\widetilde{\mathbf D} -
\mathbf{C} ) = 0.
\end{equation}
By combining~(\ref{relation1}) and~(\ref{relation2}), we see that
$\widetilde{\beta}-\beta'= - \sigma\eta_1 \cdot(\widetilde
{\mathbf C} \cdot\mathbf D/\widetilde{\mathbf D}-\mathbf C )$,
and hence $\widetilde{\beta} = \beta' - \sigma\eta_1 \cdot
(\widetilde{\mathbf C} \cdot\mathbf D/\widetilde{\mathbf D}-\mathbf
C )
$.
Hence the sets $Q_t^{(J)}(Z_{1: t}^{(J)})$ are easily updated to
$Q_t^{(J)}(\widetilde{Z}_{1: t}^{(J)})$. This procedure is repeated for
each random effect.

\section{Prior distribution selection} \label{bayesappendix}

The R package ``rjags'' was used to implement the Bayesian methods used
in the simulation study and applications. \citet{Gelman2006} suggests
using a uniform prior [i.e., $U(0,a)$] on the standard deviation
parameters when there are at least 5 groups and explains that fewer
than 3 groups results in an improper posterior distribution.
Calibration is necessary in selecting the parameter $a$ in the prior
distribution; we use 1.5 and 3 times the range of the data (per the
recommendation in Gelman [(\citeyear{Gelman2006}), page 528] to use
a
value that is
``high but not off the scale''), which appears reasonable when reviewing
the resulting posterior distributions. In the simulation study, the
results at these scale are denoted by $\bbbf{\mathrm{BAY1}_{1.5}}$ and
$\bbbf{\mathrm{BAY1}_3}$, respectively. For example, the hierarchical
model for the two-fold nested model of~(\ref{model1}) is
\begin{eqnarray*}
Y_{ijk} &\sim& N\bigl(\mu+\alpha_i + \beta_{ij},
\sigma_{\varepsilon}^2\bigr),\qquad i=1,\ldots,I, j=1,\ldots,J_i,
k=1,\ldots,K_{ij},
\\
\alpha_i &\sim& N(0, \sigma_{\alpha}),\qquad i=1,\ldots,I,\qquad
\beta_{ij} \sim N(0, \sigma_{\beta}),\qquad j=1,\ldots,J_i,
\\
\sigma_{\alpha} &\sim& U(0, a),\qquad \sigma_{\beta} \sim U(0,
a).
\end{eqnarray*}
For the second Bayesian method, a similar hierarchical model is used.
Instead of a uniform distribution on the nonerror variance components,
a half-Cauchy distribution with scale parameter $a$ set as 1.5 or 3
times the range of the data (denoted $\bbbf{\mathrm{BAY2}_{1.5}}$ and
$\bbbf{\mathrm{BAY2}_3}$, resp., in the simulation study) is
used.

\section{Proof of theorem} \label{prooftheorem1}
The proof of the convergence of the proposed SMC algorithm follows from
ideas presented in \citet{DoucMoulines2008}. Theorem
\ref{theorem1} will follow from proving the convergence of the
generated particles after each stage of the algorithm: sampling,
resampling and alteration. The development of the particle system using
the proposed algorithm does not follow the traditional SMC algorithm as
presented in \citet{DoucMoulines2008}, Section~2. A~distinction is
seen in the formulation of the proposed weights introduced in~(\ref{weights}) and discussed below.

Using the derivation of the proposal distribution in and above (\ref
{fullsamplingdensity}), the target distribution of (\ref
{fulltargetdensity}) and noting that $\mathbf I_{\mathbf C_t}(Z_{1: t})
= \mathbf I_{\mathbf C_{t-1}}(Z_{1: t-1}) \cdot\mathbf I_{\star
}(Z_{t,1: r-1}) \cdot\mathbf I_{(m_t(Z_{1: t-1}, Z_{t,1: r-1}),
M_t(Z_{1: t-1}, Z_{t,1: r-1}))}(z_{t,r})$ [where the $\star$ in\break $\mathbf
I_{\star}(Z_{t,1: r-1})$ indicates the lack of restriction to a
specific set of values for $Z_{t,1: r-1}$], the marginal target
distribution at time $t$ is
\begin{eqnarray*}
\widehat{\pi}_{1: t-1} &=& \int\pi_{1: t}(Z_{1: t})\,dZ_t = \int
\frac{ \exp(-(Z_{1: t}^T \cdot
Z_{1: t})/2) {\mathbf I_{\mathbf C_t}}(Z_{1: t})}{\Theta_{1: t}}\,dZ_t
\\
&\propto& \pi_{1: t-1} \int\exp\bigl(-\bigl(Z_t^T
\cdot Z_t\bigr)/2\bigr)
\\
&&\hspace*{38pt}{} \times\mathbf I_{\star}(Z_{t,1: r-1}) \mathbf I_{(m_t(Z_{1: t-1},
Z_{t,1: r-1}), M_t(Z_{1: t-1},
Z_{t,1: r-1}))}(z_{t,r})\,d
Z_t
\\
&\propto& \pi_{1: t-1} \cdot\bigl(\Phi\bigl(M_t(Z_{1: t-1},
Z_{t,1: r-1})\bigr) - \Phi\bigl(m_t(Z_{1: t-1},
Z_{t,1: r-1})\bigr)\bigr),
\end{eqnarray*}
where $F$ and $\Phi$ are the standard Cauchy and standard normal
cumulative distribution functions, respectively, and $\Theta_{1: t}$ is
the normalization factor at time $t$. It then follows that the
conditional target distribution at time $t$ is
\begin{eqnarray*}
\hspace*{-4pt}&&\pi_{t|1: t-1} \\
\hspace*{-4pt}&&\qquad= \pi
_{1: t}(Z_{1: t})/\widehat{\pi}_{1: t}(Z_{1: t})\\
\hspace*{-4pt}&&\qquad\propto
{\exp\bigl(-(Z_t^T \cdot Z_t)/2\bigr)\cdot\mathbf I_{\star
}(Z_{t,1: r-1}) \cdot\mathbf I_{(m_t(Z_{1: t-1}, Z_{t,1: r-1}),
M_t(Z_{1: t-1},\mathbf Z_{t,1: r-1}))}(z_{t,r})}\\
\hspace*{-4pt}&&\qquad\quad{}/\bigl({\Phi\bigl(M_t(Z_{1: t-1},
Z_{t,1: r-1})\bigr) - \Phi\bigl(m_t(Z_{1: t-1}, Z_{t,1: r-1})\bigr)}\bigr).
\end{eqnarray*}
Finally, following the notation just below~(\ref{weights}), the
derivation of the weights at time $t$ is $W_{1: t}= \frac{\pi
_{1: t}}{\widetilde{\pi}_{1: t}} = \frac{\pi_{t|1: t-1} \cdot\widehat
{\pi}_{1: t-1}}{\widetilde{\pi}_{t|1: t-1} \cdot\widetilde{\pi
}_{1: t-1}} \propto
\exp(-z_{t,r}^2/2)(1+z_{t,r}^2)(F(M_t(Z_{1: t-1}$, $Z_{t,1: r-1})) -
F(m_t(Z_{1: t-1}, Z_{t,1: r-1}))) \cdot\frac{\pi_{1: t-1}}{\widetilde
{\pi}_{1: t-1}} \propto
W_t \cdot W_{1: t-1}$.
This proof will use the above formulation and the following notation
and definition.

A particle\vspace*{1pt} system is defined as $\{Z_{1: t}^{(J)}, W_{1: t}^{(J)}\}
_{J=1}^N$ with $Z_{1: t}^{(J)}$ sampled from the proposal distribution
$\widetilde{\pi}_{1: t}$ as defined in~(\ref{fullsamplingdensity})
targeting\vspace*{1pt} probability measure $\pi_{1: t}$ on $ (\Theta_{1: t},
\mathcal B(\Theta_{1: t}) )$, and un-normalized weights
$W_{1: t}^{(J)}$ as defined in~(\ref{weights}). Let $\widetilde{\pi
}_{t|1: t-1}$ be the marginal proposal density at time $t$ as defined
above equation~(\ref{fullsamplingdensity}), which follows a Cauchy
distribution truncated to the region $R_t \triangleq(m_t(Z_{1: t-1},
Z_{t,1: r-1}), M_t(Z_{1: t-1}, Z_{t,1: r-1}))$ defined by previously
sampled particles. For notational convenience, let $\Omega_t = \sum
_{J=1}^N W_{1: t}^{(J)}$. Define two sigma-fields $\mathcal{F}_0
\triangleq\sigma(\{Z_{1: t}^{(J)}\}_{J=1}^N,(a,b]_{1: t})$ and
$\widetilde{\mathcal{F}}_J \triangleq\mathcal{F}_0 \vee\sigma(\{
\widetilde{Z}_{1: t}^{(K)}\}_{1\leq K \leq J},\break(a,b]_{1: t})$, for $J=1,
\ldots, N$. Finally, we define proper set $B_t$
[\citet{DoucMoulines2008}, Section 2.1] where
$B_t \triangleq\{f \in L^1(\Theta_{1: t}, \pi_{1: t}), F(\cdot, |f|)
\in B_{t-1}\}$ and\break
$F(Z_{1: t-1}, f) = \int f(Z_{1: t-1}, Z_t) \mathbf I_{\star}(
Z_{t,1: r-1}) \mathbf I_{R_t}(z_{t,r})(\Phi(M_t(Z_{1: t-1},Z_{t,1:
r-1}))
-\break \Phi(m_t(Z_{1: t-1}, Z_{t,1: r-1})))\pi_{t|1: t-1}(dZ_t)$.
%
%
\begin{definition}
Following Definition 1 of \citet{DoucMoulines2008}, a weighted sample
$\{Z_{1: t}^{(J)}, W_{1: t}^{(J)}\}_{J=1,\ldots,N}$ is
\textit{consistent} for the probability measure $\pi_{1: t}$ and the proper
set $B_{t}$ if, for any $f \in B_{t}$, $\Omega_t^{-1} \sum_{J=1}^N
W_{1: t}^{(J)} f(Z_{1: t}^{(J)}) \stackrel{P}{\longrightarrow} \int
f(Z_{1: t}) \pi_{1: t}(dZ_{1: t}) \triangleq\pi_{1: t}(f)$,
and $\Omega_t^{-1} \max_{J=1}^NW_{1: t}^{(J)} \stackrel
{P}{\longrightarrow} 0$.
\end{definition}

Two additional conditions on the set $B_t$ will be required to
guarantee consistency for the particle system after the alteration
step. Let $f \in B_t$, and $\Pi$ be the projection matrix onto the
null space defined in the description of the resampling step of the
algorithm found in Appendix~\ref{appresampling}. $E[f(\widetilde
{Z}_{1: t}^{(J)})\mid\mathcal{\widetilde{F}}_{J-1}] = \int f (\eta_2
\cdot\widetilde{C}+\widetilde{D} \frac{(Z_{1: t}^{(J)} - \Pi
Z_{1: t}^{(J)})}{\|Z_{1: t}^{(J)} - \Pi Z_{1: t}^{(J)}\|} ) \,d\pi
_{\widetilde{C},\widetilde{D}}= \int f(\eta_2 \cdot\widetilde
{C}+\widetilde{D} \tau_{1: t}^{(J)})\,d\pi_{\widetilde{C},\widetilde
{D}} \triangleq h_f(Z_{1: t}^{(J)})$, where $\widetilde{C}$ and
$\widetilde{D}$ are as defined above~(\ref{generalresampling}). For
$h_f$ to be in $B_t$, $f$ must be selected so that the following two
conditions hold for any direction $\tau_{1: t}$:
%
%
\begin{eqnarray}
\label{cond1}
\int\bigl|h_f(\widetilde{Z}_{1: t})\bigr|\,d\pi_{1: t} &=& \int
\biggl|\int f(\eta_2 \cdot\widetilde{C}+\widetilde{D}
\tau_{1: t})\,d\pi_{\widetilde
{C},\widetilde{D}} \biggr|\,d\pi_{1: t} < \infty,
\\
\label{cond2}
F(\widetilde{Z}_{1: t-1},h_f) &=& \int\biggl( \int f\bigl(
\eta_2 \cdot\widetilde{C} + \widetilde{D} \tau_{1: t}'
\bigr)\,d\pi_{\widetilde
{C},\widetilde{D}} \biggr) \cdot\mathbf I_{C_t}(Z_{1: t})
\nonumber\\[-8pt]\\[-8pt]
&&\hspace*{7.3pt}{}\times\bigl(\Phi(M_t) - \Phi(m_t) \bigr)
\pi_{t\mid1: t-1}(dZ_t) < \infty,\nonumber
\end{eqnarray}
where $\tau_{1: t}' = \frac{([Z_{1: t-1}, Z_t] - \Pi[Z_{1: t-1},
Z_t])}{\|[Z_{1: t-1}, Z_t] - \Pi[Z_{1: t-1}, Z_t]\|}$. Let\vspace*{1pt}
$\widetilde {B}_t$ be the set of $f \in B_t$ such that~(\ref{cond1})
and (\ref {cond2}) hold. Then, $\widetilde{B}_t \subset B_t$, and we
replace $B_t$ with $\widetilde{B}_t$ in the definition of $B_{t+1}$.
Finally, since all bounded functions satisfy~(\ref{cond1}) and~(\ref{cond2}), $\widetilde{B}_t$ is nonempty.

The goal is to show the particle system generated from the presented
algorithm is consistent. This requires the particle system after
sampling and reweighting to be consistent, after resampling to be
consistent and after the alteration to be consistent.

After noting (i) for $f \in B_t$, $E(W_{1: t}^{(J)} f(Z_{1: t}^{(J)})
\mid\mathcal F_{J-1}) =
W_{1: t-1}^{(J)}E(W_t^{(J)}f(Z_{1: t}^{(J)})\mid\mathcal F_{J-1}) =
W_{1: t-1}^{(J)} \cdot F(Z_{1: t-1},f)$
for $J = 1, \ldots, N$, with $F$ is defined above\break and~$\mathcal
{F}_{J-1} \triangleq\sigma(\{Z_{1: t-1}^{(J)}\}_{J=1}^N,(a,b]_{1: t})
\vee\sigma(\{Z_{1: t}^{(K)}\}_{1\leq K \leq J-1},(a,b]_{1: t})$,
and\break
(ii) $W_t\widetilde{\pi}_{t\mid1: t-1} = (\Phi(M_t) - \Phi
(m_t) )\pi_{t|1: t-1}$, consistency after sampling and
reweighting closely follows the proof of Theorem 1 in \citet
{DoucMoulines2008}. Consistency after resampling follows directly from
Theorem 3 in \citet{DoucMoulines2008} and consistency after the
alteration step is addressed below in Lem\-ma~\ref{lemmaalteration}.
%
%
%
\begin{lemma}[(Alteration)] \label{lemmaalteration}
Assuming the uniformly weighted sample\break $\{Z_{1:
t}^{(J)},1\}
_{J=1}^N$ is consistent for $(\pi_{1: t},B_t)$, then the altered
uniformly weighted sample $\{\widetilde{Z}_{1: t}^{(J)},1\}_{J=1}^N$ is
consistent for $(\pi_{1: t},\widetilde{B}_t)$.
\end{lemma}
\begin{pf}
Note that the $\{\widetilde{Z}_{1: t}^{(J)},1\}_{J=1}^N$ is the
\textit{altered} particle system, while $\{Z_{1: t}^{(J)},1\}_{J=1}^N$
are the resampled particles. We note that $h_f$ is a function of $Z_{1:
t}$ because $\tau_{1: t}$ is a function of $Z_{1: t}$, and, at times,
it will be necessary to write $\tau_{1: t}^{(J)} = \tau_{1: t}(Z_{1:
t}^{(J)}$). Recall that $C$ and $D$ are defined by a decomposition of
the original particle selected to be resampled and are independent by
design. The $\widetilde{C}$ and $\widetilde{D}$ are the random
variables to be resampled according to the target distributions of $C$
and $D$ with the $\tau_{1: t}^{(J)}$ considered fixed so that
$\widetilde{Z}_{1: t} = \eta_2 \cdot\widetilde{C} +
\widetilde{D}\tau_{1: t}$.

The lemma will follow once we show
\begin{eqnarray*}
&& N^{-1}\sum_{J=1}^NE\bigl[f
\bigl(\widetilde{Z}_{1: t}^{(J)}\bigr)\mid\mathcal{\widetilde
{F}}_{J-1}\bigr] = N^{-1} \sum_{J=1}^N
h_f\bigl(Z_{1: t}^{(J)}\bigr)\\
&&\quad\longrightarrow\quad \int h_f(Z_{1: t}) \pi_{1: t}(dZ_{1: t})
=
\int f(Z_{1: t}) \pi_{1: t}(dZ_{1: t}).
\end{eqnarray*}
This is because trivially $E[f(\widetilde{Z}_{1: t}^{(J)})\mid\mathcal
{\widetilde{F}}_{J-1}] = f(\widetilde{Z}_{1: t}^{(J)})$, so all that
is needed is for:

\mbox{}\hphantom{i}(i) $N^{-1} \sum_{J=1}^N h_f(Z_{1: t}^{(J)})
\longrightarrow\int h_f(Z_{1: t}) \pi_{1: t}(dZ_{1: t})$ and

(ii) $\int
h_f(Z_{1: t}) \,d\pi_{1: t}(Z_{1: t})= \int f(Z_{1: t}) \pi_{1:
t}(dZ_{1: t})$.

Point (i) follows from~(\ref{cond1}),~(\ref{cond2}), and because $f
\!\in\!\widetilde{B}_t$ so that $h_f \!\in\! B_t$.~Now we only need to show
point (ii) that $\int h_f(Z_{1: t}) \pi_{1: t}(dZ_{1: t})= \int
f(Z_{1: t}) \pi_{1: t}(dZ_{1: t})
$. This holds because
\begin{eqnarray*}
\int h_f(Z_{1: t}) \,d\pi_{1: t}
&=& \int
h_f^{*}(\tau_{1: t}) \,d\pi_{\tau}\\
&=& \int
\biggl(\int f(\eta_2 \cdot\widetilde{C} + \widetilde{D}
\tau_{1: t})\,d\pi_{\widetilde{C},\widetilde{D}} \biggr) \,d\pi_{\tau}
\\
&=& \int\biggl(\int f(\eta_2 \cdot C+D\tau_{1: t})\,d
\pi_{C,D} \biggr) \,d\pi_{\tau}  \\
&=& \int f( \eta_2
\cdot C + D\tau_{1: t})\,d\pi_{C,D} \times d\pi_{\tau}
\\
&=& \int f(Z_{1: t}) \,d\pi_{1: t} = E_{\pi_{1: t}}
\bigl[f(Z_{1: t})\bigr],
\end{eqnarray*}
where $h_f^{*}(\tau_{1: t}) = h_f(Z_{1: t})$, the equality from line one
to line three follows because $\tau(Z_{1: t}) = \tau(\widetilde
{Z}_{1: t})$ and the equality\vspace*{1pt} in the third and fourth lines follows by Fubini's
theorem because $\widetilde{C}$ and $\widetilde{D}$ are independent
of $\tau$.
\end{pf}
\end{appendix}

\section*{Acknowledgment}

The authors would like to thank the reviewers and Professor Hari Iyer
for their many helpful comments.

\begin{supplement}
\stitle{Additional simulation results}
\slink[doi]{10.1214/12-AOS1030SUPP} 
\sdatatype{.pdf}
\sfilename{aos1030\_supp.pdf}
\sdescription{The asymptotic stability of the algorithm, with respect
to the sample size and the particle sample size, was tested, and the
simulation results are included in this document. The raw results for
the simulation study in Section~\ref{simulationstudy} are also
displayed, along with additional summary figures.}
\end{supplement}


\printaddresses

\end{document}